\documentclass[a4paper,,11pt]{article}
\usepackage{fullpage}
\usepackage{graphics,amsmath,amssymb,amsthm,accents,epic}
\usepackage{mathrsfs}
\usepackage{cite}
\usepackage[top=1 in,bottom=1 in,left=1.0 in,right=1.0 in]{geometry}
\usepackage{bm}
\usepackage{subfigure}
\usepackage{latexsym}
\usepackage{xcolor}
\usepackage[colorlinks,linkcolor=black,anchorcolor=blue,citecolor=blue]{hyperref}
\usepackage{amsfonts}
\usepackage{mathtools}
\usepackage{multirow}
\usepackage{caption}
\usepackage{ulem}
\usepackage{titlesec}

\numberwithin{equation}{section}

\def \c#1{\accentset{\circ}{#1}}
\newtheorem{prop}{Proposition}[section]

\newtheorem{remark}{Remark}[section]

\newcommand{\wh}{\widehat}
\newcommand{\wt}{\widetilde}

\newcommand{\bM}{\boldsymbol{M}}
\newcommand{\bL}{\boldsymbol{L}}
\newcommand{\br}{\boldsymbol{r}}
\newcommand{\bR}{\boldsymbol{R}}
\newcommand{\ba}{\boldsymbol{a}}
\newcommand{\bC}{\boldsymbol{C}}
\newcommand{\bK}{\boldsymbol{K}}
\newcommand{\bS}{\boldsymbol{S}}
\newcommand{\bT}{\boldsymbol{T}}
\newcommand{\bV}{\boldsymbol{V}}
\newcommand{\bW}{\boldsymbol{W}}
\newcommand{\bu}{\boldsymbol{u}}
\newcommand{\bI}{\boldsymbol{I}}
\newcommand{\bv}{\boldsymbol{v}}
\newcommand{\bw}{\boldsymbol{w}}
\newcommand{\bG}{\boldsymbol{G}}
\newcommand{\bbD}{\boldsymbol{\mathcal{D}}}
\newcommand{\bbF}{\boldsymbol{\mathcal{F}}}
\newcommand{\be}{\boldsymbol{e}}
\newcommand{\bH}{\boldsymbol{H}}
\newcommand{\bJ}{\boldsymbol{J}}
\newcommand{\bF}{\boldsymbol{F}}

\newcommand{\bs}{\boldsymbol{s}}

\newcommand{\bPhi}{\boldsymbol{\Phi}}
\newcommand{\bmaU}{\boldsymbol{\mathcal{U}}}
\newcommand{\bmaV}{\boldsymbol{\mathcal{V}}}

\newcommand{\nn}{\nonumber}
\newcommand{\st}{\hbox{\tiny\it{T}}}

\title{Discretization of the Mikhailov model}	

\author{Song-lin Zhao$^{1}$,~~Xiao-gang Mu$^{1}$,~~Da-jun Zhang$^{2,3}$\footnote{
Corresponding Author: djzhang@staff.shu.edu.cn}\\
\small{$^{1}$School of Mathematical Sciences, Zhejiang University of Technology, Hangzhou 310023,  China}\\
\small{$^{2}$Department of Mathematics, Shanghai University, Shanghai 200444, China}\\
\small{$^{3}$Newtouch Center for Mathematics of Shanghai University, Shanghai 200444, China}}

\begin{document}
	
\maketitle

\begin{abstract}

In this paper the Mikhailov model is discretized by means of the Cauchy matrix approach.
A pair of discrete Miura transformations are constructed.
The discrete Mikhailov model is a coupled system,
in which one equation comes from the compatibility of the two Miura transformations,
the other is transformed from the discrete negative order
Ablowitz-Kaup-Newell-Segur system by using the Miura transformations.
Explicit solutions, including solitons and multiple-pole solutions, are presented
via two Cauchy matrix schemes respectively, namely, the Ablowitz-Kaup-Newell-Segur type
and the Kadomtsev-Petviashvili type.
By straight continuum limits, semi-discrete and continuous Mikhailov models
together with their Cauchy matrix structures and solutions are recovered.

\begin{description}
\item[Keywords:] Mikhailov model, Cauchy matrix, discretization, Miura transformation, solution
\item[Mathematics Subject Classification:] 37K60
\end{description}

\end{abstract}

%\tableofcontents

\section{Introduction}\label{sec-1}

The Mikhailov model is a relativistically invariant coupled system
(\cite{GIK-TMP-1980}, eq.(4.21))
\begin{subequations}
\label{MM}
\begin{align}
& u_{xt}+u-2\mathrm{i}uvu_x=0, \\
& v_{xt}+v+2\mathrm{i}uvv_x=0,
\end{align}
\end{subequations}
where $\mathrm{i}$ is the imaginary unit.
It was found by Mikhailov (cf.\cite{GIK-TMP-1980}) in the research of the
massive Thirring model (in light-cone coordinates) \cite{M-JETPL-1976}
\begin{subequations}\label{MTM-1}
\begin{align}
& \mathrm{i} \chi_{1,x}-\frac{m}{2}\chi_2-g |\chi_2|^2\chi_1=0, \label{MTM-1a}\\
& \mathrm{i}\chi_{2,t}-\frac{m}{2} \chi_1-g|\chi_1|^2\chi_2=0, \label{MTM-1b}
\end{align}
\end{subequations}
where $|\chi_i|^2=\chi_i\chi_i^*$ and $*$ stands for complex conjugate.
This model describes interaction of two chiral components of a two-dimensional fermion field
$\chi=(\chi_1,\chi_2)$, where
$m$ is the mass parameter of the  fermion and $g$ is the parameter for the coupling constant
 controlling the self-interaction.
In fact, the Mikhailov model \eqref{MM} allows a reduction (via $v=u^*$)
\begin{equation}\label{FL}
u_{xt}+u-2\mathrm{i}|u|^2 u_x=0,
\end{equation}
which provides a solution to \eqref{MTM-1} (for $m=2, g=1$) by defining \cite{GIK-TMP-1980,KN-LANC-1977}
\begin{equation}\label{tran-M-MTM}
\chi_2 = u_x^* e^{-\mathrm{i}\beta}, \quad \chi_1=-\mathrm{i} u^*e^{-\mathrm{i}\beta}, \quad
 \beta=\int^{x}_{-\infty}|u_x|^2\mathrm{d}x.
\end{equation}
It is Mikhailov who first presented a Lax pair for the massive Thirring model \cite{M-JETPL-1976},
from which the Kaup-Newell (KN) spectral problem followed \cite{KN-LANC-1977,KN-JMP-1978}.
It turns out that the Mikhailov model is the first equation in the negative order
potential KN hierarchy (see, e.g.\cite{CY-JPA-2008}, eq.(2.13)),
and its Lax pair reads (e.g.\cite{CY-JPA-2008,FGZ-JPA-2013})
\begin{subequations}
\begin{align}
& \bPhi_x=\bmaU \bPhi, \quad \bPhi_t=\bmaV \bPhi,\\
& \bmaU=\left(\begin{array}{cc}
\frac{\mathrm{i}}{2}\lambda^2 & \lambda u_x \\
\lambda v_x & -\frac{\mathrm{i}}{2}\lambda^2
\end{array}
\right), \quad \bmaV=\mathrm{i}\begin{pmatrix}
	\frac{1}{2}\lambda^{-2}+uv & -\lambda^{-1}u\\
	\lambda^{-1}v & -(\frac{1}{2}\lambda^{-2}+uv)
\end{pmatrix},
\end{align}
\end{subequations}
where $\bPhi=(\phi_1,\phi_2)^{\st}$ and $\lambda$ is the spectral parameter.
It is also notable that the reduced Mikhailov model \eqref{FL}
was rederived later
%(with some transformations, see \cite{Lenells-SAPM-2009})
as a generalization of the nonlinear Schr\"odinger (NLS) equation
using its bi-Hamiltonian structure by Fokas and Lenells \cite{F-PD-1995,LF-Non-2009},
as well as derived from a physics context of modeling  propagation of nonlinear pulses in monomode optical fibers
by Lenells \cite{Lenells-SAPM-2009}.
Therefore equation \eqref{FL} is also known as the Fokas-Lenells (FL) equation,
which is a reduction of the Mikhailov model \eqref{FL},
and then can be studied from the later, e.g. \cite{LWZ-SAPM-2022}.

Recently, the Mikhailov model \eqref{FL}  was reformulated
from the Cauchy matrix structure of the self-dual Yang-Mills (SDYM) equations \cite{LLZ-SAPM-2025}.
The Cauchy matrix approach was first systematically introduced in \cite{NAH-JPA-2009}
to construct solutions of  discrete integrable equations of the
Korteweg-de Vries (KdV) type.
Later it was developed to more general cases in \cite{ZZ-SAPM-2013,XZZ-JNMP-2014},
and recently, extended to the study of the SDYM equations \cite{LQYZ-SAPM-2022,LQZ-PD-2023}.
It is not only a direct method to construct and study integrable systems,
but also a possible way to discretize integrable equations.
The key step in the discretization is to introduce proper discrete dispersion relations to replace the
continuous ones in the Cauchy matrix scheme of the original continuous integrable equations.
In such a way, for example, a negative order Ablowitz-Kaup-Newell-Segur (AKNS$(-1)$) system,
the modified Korteweg-de Vries-sine Gordon (mKdV-sG) equation
and $(2+1)$-dimensional Toda equation
have been discretized in \cite{Z-JNMP-2016}, \cite{CWZ-TMP-2023}
and \cite{SLZZZ}, respectively,
by means of the Cauchy matrix approach.
More recently, it  was extended to the study of $q$-difference equations \cite{YL-MPL-2025}.

The purpose of this paper is to discretize the Mikhailov model \eqref{MM}.
However, this will be much more complicated than the work for the AKNS$(-1)$ system in \cite{Z-JNMP-2016}
and the mKdV-sG equation in \cite{CWZ-TMP-2023}.
In fact, in the Cauchy matrix formulation of the AKNS$(-1)$ system, the final involved function
is $\bS^{(0,0)}$ (see \eqref{Sij} for definition), which is a $2\times 2$ matrix
and the simplest variable in the  Cauchy matrix scheme of the AKNS system;
the discretization of the mKdV-sG equation can be done in the Cauchy matrix scheme of the KdV family
(cf.\cite{CWZ-MMAS-2023,CWZ-TMP-2023}),
but it is complicated in combining two discrete dispersion relations together in discrete sense
(see (4.14) in \cite{CWZ-TMP-2023} in comparison with $e^{kx+(k^3+1/k)t}$ for the continuous case).
For the Mikhailov model \eqref{MM}, its formulation can be done from
the  Cauchy matrix scheme of the AKNS system \cite{LLZ-SAPM-2025}
but the elements in both $\bS^{(0,0)}$ and $\bS^{(-1,0)}$ are involved:
for the functions  $u$ and $v$ in \eqref{MM}, one is defined by an element in $\bS^{(0,0)}$,
while the other is formulated by two elements in $\bS^{(-1,0)}$.
More than that, there is a  Miura transformation (\cite{LLZ-SAPM-2025}, eq.(2.13))
to link the Mikhailov model and the  AKNS$(-1)$ system,
which comes from the Miura transformation between the $J$-matrix formulation
and $K$-matrix formulation of the SDYM equations
and plays a crucial role in the derivation of \eqref{MM}.
In the discretization of \eqref{MM},
not only need we introduce proper formulation from $\bS^{(0,0)}$ and $\bS^{(-1,0)}$
and proper discrete plane wave factors,
but also we have to find out the discrete Miura type links.
All these are the difficulties that we have to overcome in the discretization of \eqref{MM}
and also make the discretization special.

In Ref.\cite{LLZ-SAPM-2025}, the Mikhailov model \eqref{MM} was formulated
as a reduction of the SDYM equations.
Note that solutions of the SDYM equations can be constructed from
Cauchy matrix structures of
the (matrix) AKNS  and the Kadomtsev-Petviashvili (KP) hierarchies \cite{LQZ-PD-2023}.
These structures are helpful in the formulation of the solutions of the Mikhailov model \eqref{MM}.
It has been shown in \cite{LLZ-SAPM-2025} that those solutions can be realized
from two types of  Cauchy matrix schemes, namely, the AKNS type and the KP type.
In the discretization, we will also consider these two types of  Cauchy matrix schemes.

The paper is organized as follows.
In the next section, the discretization of the Mikhailov model
is implemented from the Cauchy matrix approach,
where we first present the discrete AKNS$(-1)$ system,
then derive a set of discrete Miura transformations (see \eqref{Miura-h} and \eqref{Miura-t}),
and finally obtain a discrete Mikhailov model  (see eq.\eqref{dcFL-1}).
After that, we  give explicit solutions of the discrete Mikhailov model in Sec.\ref{sec-3}.
Then in Sec.\ref{sec-4} we show straight continuum limits which
recover the semi-discrete and continuous Mikhailov models
and also recover the corresponding  two types of  Cauchy matrix schemes.
Concluding remarks are given in Sec.\ref{sec-5}.
There is an appendix where we list the unreduced discrete NLS system,
and two unreduced discrete derivative NLS systems of the KN type
and Chen-Lee-Liu type, respectively.

\section{Discretization of the Mikhailov model}\label{sec-2}

The Cauchy matrix approach is in principle starting from a Sylvester equation together with
properly defined plane wave vectors and a set of master functions.
In the following we will show how this approach works in the discretization of integrable equations
as well as in the study of integrable systems.

\subsection{Cauchy matrix schemes: KP-type and AKNS-type}\label{sec-2-1}

We start from the following Sylvester equation
\begin{equation}
\label{SE}
  \bK\bM-\bM\bL=\br\bs^{\st},
\end{equation}
where $\bK,\bL$ are constant matrices in $\mathbb{C}^{N\times N}$,
$\bM\in \mathbb{C}^{N\times N}[n,m]$,
$\br, \bs \in \mathbb{C}^{N\times 2}[n,m]$ are matrix functions of $(n,m)\in \mathbb{Z}^2$.
$\br$ and $\bs$ are composed by plane wave vectors which will be specified later.
Introduce a set of  functions
\begin{equation}
\label{Sij}
\bS^{(i,j)}=\bs^{\st}\bL^{j}(\bC+\bM)^{-1}\bK^{i}\br, ~~~ i,j\in\mathbb{Z}, 
\end{equation}
where $\bC\in\mathbb{C}^{N\times N}$ is a matrix
which is chosen such that $\bC+\bM$ is invertible.
Note that each $\bS^{(i,j)}$ is a $2\times 2$ matrix.
We call $\{\bS^{(i,j)}\}$ master functions because they are the main objects that
are used to characterize integrable equations  in the Cauchy matrix approach.
They are connected to each other by some algebraic relations when
$\bK, \bL, \bC$ satisfy certain relation.

\begin{prop}\label{prop-Sij-rec}\cite{HSZ-TMP-2025,LQZ-PD-2023}
For the master functions $\{\bS^{(i,j)}\}$ defined in \eqref{Sij} with $\bM, \bK, \bL, \br$ and $\bs$
satisfying the Sylvester equation \eqref{SE},
the following recurrence relations hold:
\begin{subequations}
\label{Sij-re=1}
	\begin{align}
			\label{Sij-re=1-po}
		\bS^{(i,j+1)}=&\bS^{(i+1,j)}-\bS^{(0,j)}\bS^{(i,0)},\\
			\label{Sij-re=1-ne}
		\bS^{(i,j-1)}=&\bS^{(i-1,j)}+\bS^{(-1,j)}\bS^{(i,-1)},
	\end{align}
\end{subequations}
provided that $\bK, \bL$ and $\bC$ satisfy
\begin{align}\label{kc-cl}
		\bK\bC-\bC\bL=\bm 0.
\end{align}
\end{prop}

These recurrence relations  will be used in our derivation.
The proof for \eqref{Sij-re=1-po} can be found in Proposition 2 in \cite{LQZ-PD-2023}
while \eqref{Sij-re=1-ne} was proved in Proposition 4 in \cite{HSZ-TMP-2025}.

As we pointed out in \cite{LQZ-PD-2023}, the constraint \eqref{kc-cl} gives rise to
two Cauchy matrix schemes, namely, the KP-type and the AKNS-type. Let us elaborate them in the following.
\begin{itemize}
\item
KP-type:  For given $\bK$ and $\bL$, if we suppose they do not share any common eigenvalues,
it follows that \eqref{kc-cl} has a unique solution $\bC=\bm 0$ (see  \cite{Sylvester-1884}).
In this case, $\bM$ is also uniquely determined by the Sylvester equation  \eqref{SE}
for given $\bK, \bL,  \br$ and $\bs$, and $\{\bS^{(i,j)}\}$ are reformulated as
\begin{equation}\label{Sij-KP}
\bS^{(i,j)}=\bs^{\st}\bL^{j}\bM^{-1}\bK^{i}\br.
\end{equation}
This provides a  Cauchy matrix scheme (which we call the KP-type) to characterize (matrix) KP system
(cf. Chapter 9.7 in \cite{HJN-book-2016}
and \cite{FZ-S-2022} for the scalar case.)
\item
AKNS-type: When $\bK=\bL$, it requires $\bK$ and $\bC$ commute. In this case, $\bC$  can be normalized to
be the identity matrix $\bI$ (see \cite{LQZ-PD-2023})
and the uniqueness of solutions to the Sylvester equation \eqref{SE}
can be fulfilled in the following way.
Let $N=2 \mathcal{N}$ be even and introduce block matrices $\bK$, $\bM$, $\br$, $\bs$:
\begin{align}
\label{Sym-KM12rts-def}
	\bK=\begin{pmatrix}
		\bK_{1}&\\
		&\bK_{2}
	\end{pmatrix}, \quad \bM=\begin{pmatrix}
		&\bM_{1}\\
		\bM_{2}&
	\end{pmatrix}, \quad \br=\begin{pmatrix}
		\br_{1}&\\
		&\br_{2}
	\end{pmatrix}, \quad \bs=\begin{pmatrix}
		& \bs_{1} \\
		\bs_{2} &
	\end{pmatrix},
\end{align}	
where for $\iota=1,2$, $\bK_{\iota}$ and $\bM_{\iota}$ are $\mathcal{N}\times \mathcal{N}$ matrices,
$\br_{\iota}$ and  $\bs_{\iota}$ are $\mathcal{N}$-th order column vectors.
With the above setting the Sylvester equation \eqref{SE} with $\bL=\bK$
can be decoupled as
\begin{equation}\label{sym-SE12}
\bK_{1}\bM_{1}-\bM_{1}\bK_{2}=\br_{1}\bs^{\st}_{2},\quad \bK_{2}\bM_{2}-\bM_{2}\bK_{1}=\br_{2}\bs^{\st}_{1}.
\end{equation}
Thus, $\bM_1$ and $\bM_2$ can be uniquely determined respectively
if we assume $\bK_1$ and $\bK_2$ do not share eigenvalues \cite{Sylvester-1884}.
This gives rise to a  Cauchy matrix scheme to formulate the AKNS system \cite{Z-ROMP-2018}.
In this case, the master functions $\{\bS^{(i,j)}\}$ take a form
\begin{equation}\label{Sij-AKNS}
\bS^{(i,j)}=\bs^{\st}\bK^{j}(\bI+\bM)^{-1}\bK^{i}\br.
\end{equation}
\end{itemize}

We end the discussion of this subsection with the following remarks.

\begin{remark}\label{Rem-1}
In the following investigation, we always assume $\bK$ and $\bL$ do not share any common eigenvalues in the
KP-type scheme and $\bK_1$ and $\bK_2$ do not share any common eigenvalues
in the AKNS scheme.
In the discretization implemented in the next subsection we still employ the generic Sylvester equation \eqref{SE}
and assume it determines $\bM$ uniquely,
while we will back to the two special schemes in Sec.\ref{sec-3} when we present solutions
for the obtained discrete Mikhailov model.
\end{remark}

\begin{remark}\label{Rem-2}
We only need to consider the canonical forms of $\bK$ and $\bL$ because any matrices
that are similar to them lead to same $\bS^{(i,j)}$.
In fact, let
\begin{subequations}
\begin{align}
&		\bar{\bK}=\bT_1\bK\bT_1^{-1},\quad\bar{\bL}=\bT_2\bL\bT_2^{-1}, \label{trans-KL}\\
& \bar{\bM}=\bT_1\bM\bT_2^{-1},\quad\bar{\br}=\bT_1\br,
\quad\bar{\bs}=(\bT_2^{-1})^{\st}~\bs,\quad\bar{\bC}=\bT_1\bC\bT_2^{-1}, \label{trans-M}
\end{align}
\end{subequations}
where $\bT_{1}$ and $\bT_{2}$ are transform matrices. It is easy to see that
\begin{equation}
\label{SE-bar}
 \bar{\bK}\bar{\bM}-\bar{\bM}\bar{\bL}=\bar{\br}\bar{\bs}^{\st}
\end{equation}
and
\begin{equation}
\bS^{(i,j)}=\bs^{\st}\bL^{j}(\bC+\bM)^{-1}\bK^{i}\br
=\bar{\bs}^{\st}\bar{\bL}^{j}(\bar{\bC}+\bar{\bM})^{-1}\bar{\bK}^{i}\bar{\br},
\end{equation}
which indicates that  $\{\bS^{(i,j)}\}$ are invariant in terms of similarity transformations of $\bK$ and $\bL$.
\end{remark}

\subsection{Cauchy matrix approach to the discrete Mikhailov model}\label{sec-2-2}

\subsubsection{Dynamics of master functions}\label{sec-2-2-1}

In the discrete context, functions are defined on discrete independent variables $n,m,\cdots$,
and derivatives are replaced by shifts with respect to these independent variables.
Before we proceed, let us recall some conventional notations used in discrete equations.
Suppose $f(n,m)$ is a function (can be a matrix function) defined on $\mathbb{Z}^2$. We denote
\begin{equation}
f:=f(n,m),~~ \wt{f}:=f(n+1,m),~~ \wh{f}:=f(n,m+1),~~ \wh{\wt{f}}:=f(n+1,m+1).
\end{equation}

To obtain a discrete version of the Mikhailov model \eqref{MM}, apart from the Sylvester equation \eqref{SE},
i.e.
\begin{subequations}
\label{DES}
\begin{equation}
\label{SE1}
  \bK\bM-\bM\bL=\br\bs^{\st},
\end{equation}
we introduce the following discrete dispersion relations:
\begin{align}
\label{sh-r}
& p\wt{\br}=p\br+\bK\br\ba,\quad q\wh{\br}=q\br+\bK^{-1}\br\ba,\\
\label{sh-s}
& p\wt{\bs}+\bL^{\st}~\wt{\bs}\ba=p\bs,\quad q\wh{\bs}+(\bL^{-1})^{\st}~\wh{\bs}\ba=q\bs,
\end{align}
\end{subequations}
where  $\ba=\mathrm{diag}(1,-1)$ is the Pauli matrix $\sigma_3$,
$p$ and $q$ serve as spacing parameter in $n$-direction and $m$-direction, respectively.

For given $\bK$ and $\bL$, one can get $\br$ and $\bs$ from \eqref{sh-r} and \eqref{sh-s}.
Note that in light of Remark \ref{Rem-1}, we assume
the Sylvester equation \eqref{SE1} has a unique solution $\bM$.
It then follows that the following evolutions for $\bM$ hold: (\cite{HSZ-TMP-2025}, Proposition 6)
\begin{subequations}
		\label{shift-M}
	\begin{align}
		\label{tshift-M}
		p(\wt{\bM}-\bM)=&\br\ba\wt{\bs}^{\st},\\
		\label{hshift-M}
		q(\bM-\wh{\bM})=&\bK^{-1}\br\ba\wh{\bs}^{\st}\bL^{-1}.
	\end{align}
	\end{subequations}
One can refer to  \cite{HSZ-TMP-2025} for the proof.

Equations \eqref{shift-M} encode all the information on the dynamics of the matrix $\bM$, 
with respect to  the discrete
variables $n$ and $m$.
Then, in light of   equation set \eqref{DES} together with the above dynamics \eqref{shift-M} for $\bM$,
one can derive the following shift relations of the master functions $\{\bS^{(i,j)}\}$:
\begin{subequations}
		\label{sh-Sij}
		\begin{align}
			\label{p-Sij-1}
			p\wt{\bS}^{(i,j)}+\ba\wt{\bS}^{(i,j+1)}
=&p\bS^{(i,j)}+\bS^{(i+1,j)}\ba-\bS^{(0,j)}\ba\wt{\bS}^{(i,0)},\\
			\label{p-Sij-2}	
p\bS^{(i,j)}-\ba\bS^{(i,j+1)}=&p\wt{\bS}^{(i,j)}-\wt{\bS}^{(i+1,j)}\ba+\wt{\bS}^{(0,j)}\ba\bS^{(i,0)},\\
			\label{q-Sij-1}
			q\wh{\bS}^{(i,j)}+\ba\wh{\bS}^{(i,j-1)}
=&q\bS^{(i,j)}+\bS^{(i-1,j)}\ba+\bS^{(-1,j)}\ba\wh{\bS}^{(i,-1)},\\
			\label{q-Sij-2}
		q\bS^{(i,j)}-\ba\bS^{(i,j-1)}=&q\wh{\bS}^{(i,j)}-\wh{\bS}^{(i-1,j)}\ba-\wh{\bS}^{(-1,j)}\ba\bS^{(i,-1)}.
		\end{align}
	\end{subequations}
Again, one can refer to \cite{HSZ-TMP-2025} (Proposition 7) for the proof of the above relations.
Note that in the proof the constraint \eqref{kc-cl} was used.

\subsubsection{Discrete AKNS$(-1)$ system}\label{sec-2-2-2}

To have a discrete  Mikhailov model, we concentrate on
the functions $\bS^{(0,0)}$, $\bS^{(-1,0)}$ and $\bS^{(0,-1)}$,
and for convenience, we introduce the following notations
\begin{align}
	\label{uvw-def}
		 &\bu=\bS^{(0,0)}=\begin{pmatrix}
			u_{1}&u_{2}\\
			u_{3}&u_{4}
		\end{pmatrix}, \quad \bv=\bI-\bS^{(-1,0)}=\begin{pmatrix}
			v_{1}&v_{2}\\
			v_{3}&v_{4}
		\end{pmatrix},\quad \bw=\bI+\bS^{(0,-1)},
	\end{align}
where $u_i, v_{i}~(i=1,2,3,4)$ are scalar functions.
Note that in this paper we employ $\bI$ to express the identity matrix
without mentioning its order, which does not make confusions.
Next we will work on the variables  $\bu, \bv$ and $\bw$.

We observe that equation \eqref{Sij-re=1-ne} with specific
index combinations $(i,j)=(0,0)$,  $(1,0)$ and $(0,1)$ yields
the following relations in terms of $\bu, \bv$ and $\bw$:
\begin{subequations}\label{vw-re}
\begin{align}
& \bv\bw=\bI, \label{vw-re-1}\\
& \bS^{(1,-1)}=\bw\bu, \label{vw-re-2}\\
& \bS^{(-1,1)}=\bu\bv. \label{vw-re-3}
\end{align}
\end{subequations}
Among them the first one indicates that $\bw$ is the inverse of $\bv$.
In addition, shift relations \eqref{q-Sij-1} and \eqref{q-Sij-2} with $i=j=0$ yield
\begin{align}
	\label{h-shift-uvw}
	\bv\ba\wh{\bw}=\wh{\bv}\ba\bw=\ba-q(\wh{\bu}-\bu),
\end{align}
which, in light of \eqref{vw-re}, gives rise to
\begin{align}
	\label{h-shift-u}
	(\ba-q(\wh{\bu}-\bu))^2 =\bI,
\end{align}
which exhibits the evolution of $\bu$ in $m$-direction.

To understand the evolution of $\bu$ in $n$-direction, we consider  \eqref{p-Sij-1}
and \eqref{p-Sij-2}.
Taking  $(i,j)=(-1,0)$ in \eqref{p-Sij-1} and \eqref{p-Sij-2},
and substituting  \eqref{vw-re-3} to replace variable  $\bS^{(-1,1)}$,
we get
\begin{subequations}	\label{t-shift-uvw1}
\begin{align}
	\label{t-shift-uv1}		
p(\wt{\bv}-\bv)=(\ba\wt{\bu}-\bu\ba)\wt{\bv}=(\ba\bu-\wt{\bu}\ba)\bv.
\end{align}
Similarly, taking $(i,j)=(0,-1)$ in \eqref{p-Sij-1} and \eqref{p-Sij-2} and using \eqref{vw-re-2}, we get
\begin{align}
	p(\wt{\bw}-\bw)=\bw(\bu\ba-\ba\wt{\bu})=\wt{\bw}(\wt{\bu}\ba-\ba\bu). \label{t-shift-uw2}		
\end{align}
\end{subequations}
Then, equation \eqref{t-shift-uv1} multiplied by $\wt{\bw}$ from right,
equation \eqref{t-shift-uw2} multiplied by $\wt{\bv}$ from left,
and making use of \eqref{vw-re-1} yield
\begin{subequations}
    \label{t-shift-uvw2}
    \begin{align}
    	\label{t-shift-uvw3}
    	p\bv\wt{\bw}=&p\bI+\bu\ba-\ba\wt{\bu},\\
    	\label{t-shift-uvw4}
    	p\wt{\bv}\bw=&p\bI+\ba\bu-\wt{\bu}\ba.
    \end{align}
\end{subequations}
Now, multiplication of these two equations and making use of relation \eqref{vw-re-1} lead  to
\begin{align}
	\label{t-shift-u}
	(p\bI+\bu\ba-\ba\wt{\bu})(p\bI +\ba\bu-\wt{\bu}\ba)=p^2\bI,
\end{align}
which constitutes the evolution of $\bu$ in $n$-direction.

Next, consider the equality
\begin{align}
	\label{vw-equa}
\wt{(\bv\ba\wh{\bw})}\wh{(\wt{\bv}\bw)}
=(\wt{\bv}\bw)(\bv\ba\wh{\bw}),
\end{align}
which follows from \eqref{vw-re-1}.
Inserting \eqref{h-shift-uvw} and \eqref{t-shift-uvw4} into it
we have an equation in terms of $\bu$:
\begin{align}
	\label{u-closed}
	(\ba-q(\wh{\wt{\bu}}-\wt{\bu}))(p\bI+\ba\wh{\bu}-\wh{\wt{\bu}}\ba)
=(p\bI+\ba\bu-\wt{\bu}\ba)(\ba-q(\wh{\bu}-\bu)).
\end{align}
Note that there is an alternative equality which is from \eqref{vw-re-1} as well:
\begin{align}
	\label{vw-equa-1}
	(\wh{\bv}\ba\bw)(\bv\wt{\bw})=\wh{(\bv\wt{\bw})}\wt{(\wh{\bv}\ba\bw)},
\end{align}
from which and \eqref{h-shift-uvw} and \eqref{t-shift-uvw3} we get
an alternative  $\bu$ equation:
\begin{align}
	\label{u-closed-1}
(p\bI+\wh{\bu}\ba-\ba\wh{\wt{\bu}})(\ba-q(\wh{\wt{\bu}}-\wt{\bu}))
=(\ba-q(\wh{\bu}-\bu))(p\bI+\bu\ba-\ba\wt{\bu}).
\end{align}
In fact, the later is an inverse of the former:
multiplying \eqref{u-closed} and \eqref{u-closed-1}
one get the identity matrix on both sides.
$\bu$ is a $2\times 2$ matrix with 4 elements $u_1, u_2, u_3$ and $u_4$.
One can make use of \eqref{h-shift-u}, \eqref{t-shift-u} to remove $u_1$ and $u_4$
from either \eqref{u-closed} or \eqref{u-closed-1},
and then get a coupled quadrilateral system for $u_2$ and $u_3$ \cite{Z-JNMP-2016}:
\begin{subequations}
	\label{nAKNS}
	\begin{align}
		\label{nAKNS1}
& q(\wh{u}_{2}-u_{2})\Psi_{1}+(\wt{u}_{2}+u_{2})\Psi_{2}
+q(\wt{u}_{2}-\wh{\wt{u}}_{2})\wh{\Psi}_{1}+(\wh{u}_{2}+\wh{\wt{u}}_{2})\wt{\Psi}_{2}=0,\\
		\label{nAKNS2}
& q(\wh{u}_{3}-u_{3})\Psi_{1}+(\wt{u}_{3}+u_{3})\Psi_{2}
+q(\wt{u}_{3}-\wh{\wt{u}}_{3})\wh{\Psi}_{1}+(\wh{u}_{3}+\wh{\wt{u}}_{3})\wt{\Psi}_{2}=0,
\end{align}
where
\begin{align}
	\label{Psi-def}
 \Psi_{1}=[p^2-(\wt{u}_{2}+u_{2})(\wt{u}_{3}+u_{3})]^{\frac{1}{2}},\quad
		\Psi_{2}=[1-q^2(\wh{u}_{2}-u_{2})(\wh{u}_{3}-u_{3})]^{\frac{1}{2}}.
\end{align}
\end{subequations}
This is known as the discrete AKNS$(-1)$ system, whose continuum limit
recovers the continuous AKNS$(-1)$ system \cite{HSZ-TMP-2025}.

\subsubsection{Miura transformations}\label{sec-2-2-3}

After the discrete AKNS$(-1)$ system which is formulated by $\bu$, we take a close look at $\bv$ and $\bw$.
First, from \eqref{h-shift-uvw} we find
$|\bv||\wh{\bw}|=|\wh{\bv}||\bw|$,
which, in light of \eqref{vw-re-1}, yields a relation
$|\bv|^2=|\wh{\bv}|^2$,
which indicates $|\bv|^2$ is independent of $m$.
In addition, from \eqref{t-shift-uv1} we know that
$|\ba\wt{\bu}\ba-\bu||\wt{\bv}|=|\bu-\ba\wt{\bu}\ba||\bv|$.
Since $\bu$ is a $2\times 2$ matrix and $\ba=\ba^{-1}$, we have $|\wt{\bv}|=|\bv|$
which means $\bv$ is irrelevant to $n$.
As a result, $|\bv|^2$ must be a constant, and so is $|\bw|^2$.
In fact, in Sec.\ref{sec-4} we can see that
in the AKNS-type and KP-type Cauchy matrix schemes, $|\bv|$ is a constant.
Thus, in the following discussions, one can normalize $\bv$ and $\bw$ by
\begin{align}
	\label{trans1}
	\bV= \frac{1}{\sqrt{|\bv|}}\bv=\begin{pmatrix}
		V_{1}&V_{2}\\
		V_{3}&V_{4}
	\end{pmatrix}, \quad \bW= \frac{1}{\sqrt{|\bw|}}\bw
\end{align}
and consider $\bV$ and $\bW$,
or in stead, still employ the $\bv$ and $\bw$ but assume  $|\bv| =1$.
We prefer the later, i.e.  assuming  $|\bv| =1$ without loss of generality.
Thus,
$\bw$ defined in \eqref{uvw-def} can be expressed as
\begin{align}
	\label{vw-def1}
\bw=\bI+\bS^{(0,-1)}=\begin{pmatrix}
		v_{4}&-v_{2}\\
		-v_{3}&v_{1}
	\end{pmatrix}.
\end{align}
As a result, equation \eqref{h-shift-uvw} can be represented in terms of $\{u_i\}$ and $\{v_i\}$:
\begin{subequations}
	\label{h-component-uv}
	\begin{align}
		\label{h-component1}
		&\wh{v}_{2}v_{3}+\wh{v}_{1}v_{4}=v_{2}\wh{v}_{3}+v_{1}\wh{v}_{4},\\
		\label{h-component3}
		&A_{1}\coloneqq1-q(\wh{u}_{1}-u_{1})=1+q(\wh{u}_{4}-u_{4})=v_{1}\wh{v}_{4}+v_{2}\wh{v}_{3},\\
		\label{h-component4}
		&A_{2}\coloneqq q(\wh{u}_{2}-u_{2})=v_{1}\wh{v}_{2}+\wh{v}_{1}v_{2},\\
		\label{h-component2}
		&A_{3}\coloneqq-q(\wh{u}_{3}-u_{3})=v_{3}\wh{v}_{4}+\wh{v}_{3}v_{4}.
	\end{align}
\end{subequations}
Note that \eqref{h-component1}  is a consequence of
$\bv\ba\wh{\bw}=\wh{\bv}\ba\bw$,
while the other three are due to $\wh{\bv}\ba\bw=\ba-q(\wh{\bu}-\bu)$.

In the next step, we introduce a new variable
\begin{equation}\label{nu}
\nu\coloneqq \mathrm{i}\frac{v_{2}}{v_{4}}.
\end{equation}
We aim to get a discrete Mikhailov model in terms of $u_3$ and $\nu$ (see equation \eqref{dcFL-1}).
To achieve that, we need to look for a set of Miura transformations to connect the discrete
AKNS$(-1)$ system \eqref{nAKNS} and \eqref{dcFL-1},
and meanwhile, we will also show that with the help of the Miura transformations,
$\Psi_1$ and $\Psi_2$ defined in \eqref{Psi-def} can be re-expressed
in some proper forms.

We revisit \eqref{h-component-uv} by substituting  $v_{2}=-\mathrm{i}v_{4}\nu$.
At first, for $A_{2}$ we have
\begin{align}
	\label{h-v5-1}
	A_{2}=-\mathrm{i}(v_{1}\wh{v}_{4}\wh{\nu}+\wh{v}_{1}v_{4}\nu).
\end{align}
Then, using \eqref{h-component1} to replace $v_{1}\wh{v}_{4}$ yields
\begin{align}\label{h-v5-2}
A_{2}=-\mathrm{i}\wh{v}_{1}v_{4}(\wh{\nu}+\nu)-v_{3}\wh{v}_{4}\wh{\nu}^{2}
+\wh{v}_{3}v_{4}\nu\wh{\nu}.
\end{align}
Further, we use \eqref{h-component2} to replace $v_{3}\wh{v}_{4}$ and we have
\begin{align}
	\label{h-v5-3}
	A_{2}=(-\mathrm{i}\wh{v}_{1}v_{4}+\wh{v}_{3}v_{4}\wh{\nu})(\wh{\nu}+\nu)-A_{3}\wh{\nu}^2.
\end{align}
Again, replacing the term $\wh{v}_{1}v_{4}$ by using $A_{1}$ \eqref{h-component3},
we arrive at
\begin{align}
	\label{h-v5-4}
	A_{2}-A_{3}\nu\wh{\nu}=-\mathrm{i}A_{1}(\wh{\nu}+\nu),
\end{align}
i.e.
\begin{align}
	\label{A1}
A_1=\mathrm{i}\frac{	A_{2}-A_{3}\nu\wh{\nu}}{\wh{\nu}+\nu}.
\end{align}
On the other hand, revisit \eqref{h-shift-u} which is a consequence of \eqref{h-shift-uvw} and \eqref{vw-re-1}.
Expand  \eqref{h-shift-u} and it turns out that the $(1,2)$ and $(2,1)$
elements agree with \eqref{h-component2},
while the $(1,1)$ and $(2,2)$ elements give rise to a same relation in term of $\{A_{i}\}$:
\begin{align}
	\label{Ai-relation}
	A_{1}^2-A_{2}A_{3}=1.
\end{align}
Thus we can insert \eqref{A1} into it and then solve the equation for solution $A_{2}$, 
which is expressed as\footnote{
In principle, there is another solution for $A_2$, which is obtained by replacing $\mathrm{i}$
by $-\mathrm{i}$ from \eqref{h-u2u3v5-1}.
However, since $\{A_i\}$ are formulated by $\{v_i\}$
which have been defined by \eqref{DES} and \eqref{Sij},
we choose the form \eqref{h-u2u3v5-1} because it agrees with the explicit solutions presented in Sec.\ref{sec-3}.
We also note that for the same reason we take $A_1=\left( 1+A_{2}A_{3}\right)^{\frac{1}{2}}$
and choose the signs in \eqref{t-u2u3v5-1} and \eqref{Psi1-ndef} as they are.
}
\begin{align}
	\label{h-u2u3v5-1}
A_{2}=-\frac{1}{2} A_{3}(\wh{\nu}^{2}+\nu^{2})
-\frac{1}{2}\mathrm{i}(\wh{\nu}+\nu)[4-A_{3}^{2}(\wh{\nu}-\nu)^2] ^{\frac{1}{2}},
\end{align}
i.e.
\begin{align}\label{Miura-h}
2(\wh{u}_{2}-u_{2})
=(\wh{u}_{3}-u_{3})(\wh{\nu}^{2}+\nu^{2})
-\mathrm{i}(\wh{\nu}+\nu)[4/q^2-(\wh{u}_{3}-u_{3})^2(\wh{\nu}-\nu)^2] ^{\frac{1}{2}}.
\end{align}
This is one equation in the Miura transformations to connect the discrete
AKNS$(-1)$ system \eqref{nAKNS} and the discrete Mikhailove model (see equation \eqref{dcFL-1}).\footnote{
In discrete case usually one equation is not enough to compose a Miura transformation.
This has been demonstrated by many examples, e.g. Chapter 9 in \cite{HJN-book-2016}.}
In addition, recalling $\Psi_2$ defined in \eqref{Psi-def} which can be written in terms of
$A_2$ and $A_3$ as
\begin{equation}
	\label{Psi2-ndef23}
	\Psi_{2}=\left( 1+A_{2}A_{3}\right)^{\frac{1}{2}},
\end{equation}
which leads us to
\begin{align}
	\label{Psi2-ndef}
	\Psi_{2}=A_{1}=\frac{\mathrm{i}q[\wh{u}_{2}-u_{2}+(\wh{u}_{3}-u_{3})\wh{\nu}\nu]}{\wh{\nu}+\nu},
\end{align}
where \eqref{Ai-relation} has been used.
Apparently, in light of the Miura transformation \eqref{Miura-h},
$\Psi_2$ can be written as a function of $u_3$ and $\nu$.

Next, we look for another formula in the Miura transformations.
We rewrite  \eqref{t-shift-uv1} in explicit form, which yields
\begin{subequations}
	\label{t-component-uv}
	\begin{align}
		\label{t-v1}
		&p(\wt{v}_{1}-v_{1})=B_{2}v_{3}-B_{1}v_{1}=B_{2}\wt{v}_{3}+B_{1}\wt{v}_{1},\\
		\label{t-v3}
		&p(\wt{v}_{3}-v_{3})=B_{3}v_{1}-B_{1}v_{3}=B_{3}\wt{v}_{1}+B_{1}\wt{v}_{3},\\
		\label{t-v2}
		&p(\wt{v}_{2}-v_{2})=B_{2}v_{4}-B_{1}v_{2}=B_{2}\wt{v}_{4}+B_{1}\wt{v}_{2},\\
		\label{t-v4}
		&p(\wt{v}_{4}-v_{4})=B_{3}v_{2}-B_{1}v_{4}=B_{3}\wt{v}_{2}+B_{1}\wt{v}_{4},
	\end{align}
\end{subequations}
where
\begin{subequations}\label{B}
\begin{align}
& B_{1}:=\wt{u}_{1}-u_{1}=-(\wt{u}_{4}-u_{4}),\label{B1}\\
& B_{2}:=\wt{u}_{2}+u_{2}, \quad B_{3}:=-\wt{u}_{3}-u_{3}.\label{B2}
\end{align}
\end{subequations}
For convenience, we rewrite those relations in \eqref{t-component-uv}
that involve $\{B_i\}$ and $\{v_i\}$ as follows
\begin{subequations}
	\label{t-Bivi}
	\begin{align}
		\label{t-Bivi-1}
		&B_{2}(\wt{v}_{3}-v_{3})=-B_{1}(\wt{v}_{1}+v_{1}),\quad B_{3}(\wt{v}_{1}-v_{1})=-B_{1}(\wt{v}_{3}+v_{3}),\\
		\label{t-Bivi-2}
		&B_{2}(\wt{v}_{4}-v_{4})=-B_{1}(\wt{v}_{2}+v_{2}),\quad B_{3}(\wt{v}_{2}-v_{2})=-B_{1}(\wt{v}_{4}+v_{4}).
	\end{align}
\end{subequations}
Performing division on the two equations in \eqref{t-Bivi-2} and then using
 \eqref{t-v2} and \eqref{t-v4} yield
\begin{align}
	\frac{\wt{v}_{2}+v_{2}}{\wt{v}_{4}+v_{4}}=\frac{B_{2}(\wt{v}_{4}-v_{4})}{B_{3}(\wt{v}_{2}-v_{2})}
=\frac{B_{2}(B_{3}\wt{v}_{2}+B_{1}\wt{v}_{4})}{B_{3}(B_{2}\wt{v}_{4}+B_{1}\wt{v}_{2})},
\end{align}
i.e.
\begin{align}
	\label{t-Biv2v4}
	B_{1}B_{3}\wt{v}_{2}^{2}+B_{1}B_{3}v_{2}\wt{v}_{2}+B_{2}B_{3}v_{2}\wt{v}_{4}
=B_{1}B_{2}\wt{v}_{4}^{2}
+B_{1}B_{2}v_{4}\wt{v}_{4}+B_{2}B_{3}v_{4}\wt{v}_{2}.
\end{align}
Meanwhile, we can represent  \eqref{t-v2} and \eqref{t-v4} as
\begin{align}
	\label{t-Biv2v4-1}
	p\wt{v}_{2}^{2}=(p-B_{1})v_{2}\wt{v}_{2}+B_{2}\wt{v}_{2}v_{4},\quad
	p\wt{v}_{4}^{2}=(p-B_{1})v_{4}\wt{v}_{4}+B_{3}\wt{v}_{4}v_{2}.
\end{align}
Then, using them to replace $p\wt{v}_{2}^{2}$ and $p\wt{v}_{4}^{2}$ in \eqref{t-Biv2v4},
 we get
\begin{align}
	\label{t-Biv2v4-2}
	(2pB_{1}-B_{1}^{2})(B_{3}v_{2}\wt{v}_{2}-B_{2}v_{4}\wt{v}_{4})
+(B_{1}-p)B_{2}B_{3}(v_{4}\wt{v}_{2}-v_{2}\wt{v}_{4})=0,
\end{align}
which is rewritten in terms of $\{B_i\}$ and $\nu$ as
\begin{align}
	\label{t-Biv5}
	(2pB_{1}-B_{1}^{2})(B_{2}+B_{3}\nu\wt{\nu})+\mathrm{i}(B_{1}-p)B_{2}B_{3}(\wt{\nu}-\nu)=0.
\end{align}
On the other hand, expanding relation \eqref{t-shift-u} we get
\begin{align}
	\label{Bi-relation}
	2pB_{1}-B_{1}^2=-B_{2}B_{3},
\end{align}
using which \eqref{t-Biv5} is simplified as
\begin{align}
	\label{t-Biv5-1}
	B_{2}+B_{3}\nu\wt{\nu}=\mathrm{i}(p-B_{1})(\nu-\wt{\nu}).
\end{align}
Now we combine \eqref{Bi-relation} and \eqref{t-Biv5-1} together,
eliminate $B_1$ from them and get a second order algebraic  equation for $B_{2}$.
Solving it leads to the following expression
\begin{align}
	\label{t-u2u3v5-1}
B_{2}=-\frac{1}{2}B_{3}(\wt{\nu}^{2}+\nu^{2})
-\frac{1}{2}\mathrm{i}(\wt{\nu}-\nu)[4p^2 -B_{3}^{2}(\wt{\nu}+\nu)^2]^{\frac{1}{2}},
\end{align}
i.e. in light of \eqref{B2},
\begin{align}
	\label{Miura-t}
	2(\wt{u}_{2}+u_{2})=(\wt{u}_{3}+u_{3})(\wt{\nu}^{2}+\nu^{2})
-\mathrm{i}(\wt{\nu}-\nu)[4p^2-(\wt{u}_{3}+u_{3})^{2}(\wt{\nu}+\nu)^{2}]^{\frac{1}{2}}.
\end{align}
This provides the second formula in the Miura transformations.
In addition, from \eqref{Bi-relation} we have
\begin{align}
	\label{Bi-relation-1}
 (p-B_{1})^{2}=p^{2}+B_{2}B_{3},
\end{align}
and meanwhile noting that $\Psi_{1}$ given in \eqref{Psi-def} has the form
\begin{align}
	\label{Psi1-ndef0}
	\Psi_{1}=(p^{2}+B_{2}B_{3})^{\frac{1}{2}},
\end{align}
one can rewrite it as
\begin{align}
	\label{Psi1-ndef}
	\Psi_{1}=p-B_{1}=\frac{\mathrm{i}[\wt{u}_{2}+u_{2}-(\wt{u}_{3}+u_{3})\nu\wt{\nu}]}{\wt{\nu}-\nu},
\end{align}
which can be expressed in terms of only $u_3$ and $\nu$ if using the Miura transformation \eqref{Miura-t}.

\subsubsection{Discrete Mikhailov model}\label{sec-2-2-4}

Now we are at a stage to present a discrete version of the Mikhailov model.
We may substitute $\Psi_2$ from \eqref{Psi2-ndef} and $\Psi_1$ from \eqref{Psi1-ndef}
together with the two Miura transformations into equation \eqref{nAKNS2}.
In this way we convert equation \eqref{nAKNS2} into an equation which involves only $u_3$ and $\nu$,
i.e. \eqref{dcFL-1a} in the following.
On the other hand, by checking the compatibility of $u_2$ in the two Miura transformations
i.e. \eqref{Miura-h} and \eqref{Miura-t},
we get another equation for only $u_3$ and $\nu$, i.e. \eqref{dcFL-1b}.
Thus, we arrive at the following coupled quadrilateral system for
$(\mu,\nu)\coloneqq(u_{3},\mathrm{i}v_{2}/v_{4})$:
\begin{subequations}
	\label{dcFL-1}
	\begin{align}
		\label{dcFL-1a}
		& (\wh{\nu}+\wt{\nu})(\wh{\wt{\mu}}+\mu)(\wh{\wt{\mu}}+\wh{\mu}-\wt{\mu}-\mu) \nn \\
		& \qquad -\mathrm{i}\big[(\wh{\mu}-\mu)P+(\wt{\mu}+\mu)Q
+(\wt{\mu}-\wh{\wt{\mu}})\wh{P}+(\wh{\mu}+\wh{\wt{\mu}})\wt{Q}\big]=0, \\
		\label{dcFL-1b}
		& (\wt{\mu}+\wh{\mu})(\nu^{2}-\wh{\wt{\nu}}{}^{2})
+(\mu+\wh{\wt{\mu}})(\wt{\nu}^{2}-\wh{\nu}^{2})\nn \\
		& \qquad -\mathrm{i}\big[(\wt{\nu}-\nu)P
+(\wh{\nu}+\nu)Q+(\wh{\nu}-\wh{\wt{\nu}})\wh{P}
+(\wt{\nu}+\wh{\wt{\nu}})\wt{Q}\big]=0,
	\end{align}
where
\begin{align}
       \label{PQ-def}
	 P=[4p^2-(\mu+\wt{\mu})^{2}(\nu+\wt{\nu})^{2}]^{\frac{1}{2}},\quad
	      Q=[4/q^2-(\wh{\mu}-\mu)^2(\wh{\nu}-\nu)^2]^{\frac{1}{2}}.
\end{align}
\end{subequations}
This is a discrete version of the Mikhailov model \eqref{MM}, which will recover the continuous system
\eqref{MM} in continuum limits (see Sec.\ref{sec-4}).

Analogous to the above derivation for \eqref{dcFL-1},
one can also obtain another closed-form lattice equations with dependent variables
$(\mu',\nu')\coloneqq(u_2,\mathrm{i}v_{3}/v_{1})$:
\begin{subequations}
	\label{dcFL-2}
	\begin{align}
		\label{dcFL-2a}
		& (\wh{\nu}'+\wt{\nu}')(\wh{\wt{\mu}}{}'+\mu')(\wh{\wt{\mu}}{}'+\wh{\mu}'-\wt{\mu}'-\mu') \nn \\
		& \qquad +\mathrm{i}\big[(\wh{\mu}'-\mu')P'+(\wt{\mu}'+\mu')Q'+(\wt{\mu}'-\wh{\wt{\mu}}{}')\wh{P}'
+(\wh{\mu}'+\wh{\wt{\mu}}{}')\wt{Q}'\big]=0, \\
		\label{dcFL-2b}
		& (\wt{\mu}'+\wh{\mu}')(\nu'{}^{2}-\wh{\wt{\nu}}{}'{}^{2})
+(\mu'+\wh{\wt{\mu}}{}')(\wt{\nu}'{}^{2}-\wh{\nu}'{}^{2})\nn \\
		& \qquad +\mathrm{i}\big[(\wt{\nu}'-\nu')P'+(\wh{\nu}'+\nu')Q'
+(\wh{\nu}'-\wh{\wt{\nu}}{}')\wh{P}'+(\wt{\nu}'+\wh{\wt{\nu}}{}')\wt{Q}'\big]=0,
	\end{align}
where
\begin{align}
	 P'=[4p^2-(\mu'+\wt{\mu}')^{2}(\nu'+\wt{\nu}')^{2}]^{\frac{1}{2}},\quad
	      Q'=[4/q^2-(\wh{\mu}'-\mu')^2(\wh{\nu}'-\nu')^2]^{\frac{1}{2}}.
\end{align}
\end{subequations}
We skip presenting the derivation details.
It is notable that systems \eqref{dcFL-1}
and \eqref{dcFL-2} are highly similar in their expressions, only different from the sign
in front of $\mathrm{i}$.

In conclusion, we have in this section identified the construction of the discrete Mikhailov model from the
Cauchy matrix setting \eqref{DES}. We have made use of the discrete AKNS$(-1)$ system \eqref{nAKNS}
and the pair of Miura transformations consisting of \eqref{Miura-h} and \eqref{Miura-t}.
In the following section, we only present solution $(\mu,\nu):=(u_{3},\mathrm{i}v_{2}/v_{4})$
to the system \eqref{dcFL-1}.
% without loss of generality.

\section{Explicit solutions}\label{sec-3}

In this section we consider   solutions of the discrete Mikhailov model \eqref{dcFL-1}.
Following the treatment in \cite{LQZ-PD-2023} and \cite{HSZ-TMP-2025},
subject to the constraint \eqref{kc-cl}, we respectively present exact solutions
from the AKNS-type and the KP-type Cauchy matrix schemes.
Multi-soliton solutions and multiple-pole solutions in these two schemes
will be presented explicitly.
At this stage we introduce the discrete plane wave factors
\begin{subequations}
	\label{plane-def}
	\begin{align}
\label{plane-rho-def}
		&\rho_{\iota}(z)=\left( 1+(-1)^{\iota-1}\frac{z}{p}\right)^{n}
\left(1+ (-1)^{\iota-1}\frac{1}{qz}\right)^{m}\rho^{0}_{\iota}(z),\\
		&\sigma_{\iota}(z)=\left( 1+(-1)^{\iota-1}\frac{z}{p}\right)^{-n}
\left(1+ (-1)^{\iota-1}\frac{1}{qz}\right)^{-m}\sigma^{0}_{\iota}(z),
	\end{align}
\end{subequations}
where $\rho^{0}_{\iota}(z),~\sigma^{0}_{\iota}(z),~(\iota=1,2)$ are phase parameters independent of $(n,m)$.
They serve as the plane wave factors in the two schemes.

\subsection{The AKNS-type scheme}\label{sec-3-1}

As we have discussed in Sec.\ref{sec-2-1}, this case happens when $\bL=\bK$ and
$N=2\mathcal{N}$, where $\mathcal{N}$ stands for the
number of solitons.
In this case, the Sylvester equation \eqref{SE1} becomes
\begin{align}
\label{SE-sy}
\bK\bM-\bM\bK=\br\bs^{\st},
\end{align}
$\bC$ can be normalized to be $\bI$, and in light of the setting \eqref{Sym-KM12rts-def},
the system \eqref{DES} decouples into the following form:
\begin{subequations}
	\label{DES-sym-ent}
	\begin{align}
        \label{sym-SE12}
& \bK_{1}\bM_{1}-\bM_{1}\bK_{2}=\br_{1}\bs^{\st}_{2},\quad \bK_{2}\bM_{2}-\bM_{2}\bK_{1}=\br_{2}\bs^{\st}_{1}, \\
		\label{sym-shift-r1}
& p\wt{\br}_{\iota}=p\br_{\iota}+(-1)^{\iota-1}\bK_{\iota}\br_{\iota},\quad q\wh{\br}_{\iota}=q\br_{\iota}+(-1)^{\iota-1}\bK^{-1}_{\iota}\br_{\iota}, \\
		\label{sym-shift-s1}
& p\wt{\bs}_{\iota}+(-1)^{\iota}\bK^{\st}_{\iota}~\wt{\bs}_{\iota}=p\bs_{\iota},\quad q\wh{\bs}_{\iota}+(-1)^{\iota}(\bK_{\iota}^{-1})^{\st}~\wh{\bs}_{\iota}=q\bs_{\iota},
	\end{align}
\end{subequations}
with $\iota=1,2$,
where it is assumed that $\bK_1$ and $\bK_2$ do not share any eigenvalues.
The master functions  \eqref{Sij-AKNS} are thereby decomposed as
\begin{align}
	\label{sym-S-def}
		\bS^{(i,j)}_{[\text{AKNS}]}=&\bs^{\st}\bK^{j}(\bI+\bM)^{-1}\bK^{i}\br=\begin{pmatrix}
			s^{(i,j)}_{[\text{AKNS}1]}&s^{(i,j)}_{[\text{AKNS}2]}\\
			s^{(i,j)}_{[\text{AKNS}3]}&s^{(i,j)}_{[\text{AKNS}4]}
		\end{pmatrix} \nn \\
		=&\begin{pmatrix}
			-\bs^{\st}_{2}\bK_{2}^{j}\bM_{2}(\bI-\bM_{1}\bM_{2})^{-1}\bK_{1}^{i}\br_{1} & \bs^{\st}_{2}\bK_{2}^{j}(\bI-\bM_{2}\bM_{1})^{-1}\bK_{2}^{i}\br_{2}\\
			\bs^{\st}_{1}\bK_{1}^{j}(\bI-\bM_{1}\bM_{2})^{-1}\bK_{1}^{i}\br_{1} & -\bs^{\st}_{1}\bK_{1}^{j}\bM_{1}(\bI-\bM_{2}\bM_{1})^{-1}\bK_{2}^{i}\br_{2}
		\end{pmatrix}.
\end{align}
The matrix functions $\bu$ and $\bv$ defined in equation \eqref{uvw-def} are formulated as
\begin{align}
	\label{sym-uv-def}
	\bu=\bS^{(0,0)}_{[\text{AKNS}]}=\bs^{\st}(\bI+\bM)^{-1}\br,\quad\bv=\bI-\bS^{(-1,0)}_{\text{[AKNS]}}
=\bI-\bs^{\st}(\bI+\bM)^{-1}\bK^{-1}\br.
\end{align}
Using the Weinstein-Aronszajn formula and the Sylvester equation \eqref{SE-sy}, one can find
\begin{align}
	\label{sym-det-v}
		|\bv| &=\left|\bI-\bs^{\st}(\bI+\bM)^{-1}\bK^{-1}\br\right| \nn \\
       &= \left|\bI-(\bI+\bM)^{-1}\bK^{-1}\br\bs^{\st}\right| \nn \\
		&=\left|(\bI+\bM)^{-1}\bK^{-1}(\bI+\bM)\bK\right| \nn \\
		&=1,
\end{align}
which agrees with the assumption $|\bv|=1$ introduced in Sec.\ref{sec-2-2-3}.

For the solution $(\mu,\nu)$ to the  discrete Mikhailov model \eqref{dcFL-1},
their formulae are given by
\begin{align}
	\label{sym-FL-solu}
	(\mu,\nu)=\left( u_{3},\mathrm{i}\frac{v_{2}}{v_{4}}\right) =\left( \bs^{\st}_{1}(\bI-\bM_{1}\bM_{2})^{-1}\br_{1},
\frac{-\mathrm{i}\bs^{\st}_{2}(\bI-\bM_{2}\bM_{1})^{-1}\bK_{2}^{-1}\br_{2}}{1+\bs^{\st}_{1}
\bM_{1}(\bI-\bM_{2}\bM_{1})^{-1}\bK_{2}^{-1}\br_{2}}\right),
\end{align}
where the involved elements are determined by the system \eqref{DES-sym-ent}.

Next, we present the explicit formulae for solitons and multiple-pole solutions
through solving  \eqref{DES-sym-ent}.

\vskip 5pt
\noindent
\textbf{Soliton solutions}. To get the soliton solutions, we take
\begin{align}
\label{K12-def}
\bK_{1}=\text{diag}(k_{1},k_{2},\cdots,k_{\mathcal{N}}), \quad \bK_{2}=\text{diag}(l_{1},l_{2},\cdots,l_{\mathcal{N}}),
\end{align}
with $k_i\neq l_j,~(i,j=1,2,\ldots,\mathcal{N}$). Then we have
\begin{subequations}
	\label{sym-SE-solu}
	\begin{align}
		&\br_{1}=(\rho_{1}(k_{1}),\rho_{1}(k_{2}),\cdots,\rho_{1}(k_{\mathcal{N}}))^{\st}, \quad \br_{2}=(\rho_{2}(l_{1}),\rho_{2}(l_{2}),\cdots,\rho_{2}(l_{\mathcal{N}}))^{\st},\\
		&\bs_{1}=(\sigma_{2}(k_{1}),\sigma_{2}(k_{2}),\cdots,\sigma_{2}(k_{\mathcal{N}}))^{\st}, \quad \bs_{2}=(\sigma_{1}(l_{1}),\sigma_{1}(l_{2}),\cdots,\sigma_{1}(l_{\mathcal{N}}))^{\st},\\
		\label{AKNS-M1-def}
		&\bM_{1}=(M_{1,ij})_{\mathcal{N}\times \mathcal{N}},\quad M_{1,ij}=\frac{\rho_{1}(k_{i})\sigma_{1}(l_{j})}{k_{i}-l_{j}},\\
		\label{AKNS-M2-def}
		&\bM_{2}=(M_{2,ij})_{\mathcal{N}\times \mathcal{N}},\quad M_{2,ij}=\frac{\rho_{2}(l_{i})\sigma_{2}(k_{j})}{l_{i}-k_{j}}.
	\end{align}
\end{subequations}
For the sake of simplicity, we describe the matrices $\bM_{1},~\bM_{2},~\br_{\iota}$ and $\bs_{\iota}$ as follows
\begin{align}
	\label{sym-M1M2-decom}
	\bM_{1}=\bR_{1}\bG\bS_{2}, \quad \bM_{2}=-\bR_{2}\bG^{\st}\bS_{1},
\quad \br_{\iota}=\bR_{\iota}\be_{\mathcal{N}}, \quad \bs_{\iota}=\bS_{\iota} \be_{\mathcal{N}}, \quad (\iota=1,2),
\end{align}
where $\be_{\mathcal{N}}=(\underbrace{1, 1, \cdots, 1}_{\mathcal{N}\text{-dimensional}})^{\st}$ and
\begin{subequations}
	\label{RGS-def}
	\begin{align}
		\label{R-def}		&\bR_{1}=\text{diag}(\rho_{1}(k_{1}),\rho_{1}(k_{2}),\cdots,\rho_{1}(k_{\mathcal{N}})), \quad
\bR_{2}=\text{diag}(\rho_{2}(l_{1}),\rho_{2}(l_{2}),\cdots,\rho_{2}(l_{\mathcal{N}})), \\
		\label{S-def}	&\bS_{1}=\text{diag}(\sigma_{2}(k_{1}),\sigma_{2}(k_{2}),\cdots,\sigma_{2}(k_{\mathcal{N}})), \quad
\bS_{2}=\text{diag}(\sigma_{1}(l_{1}),\sigma_{1}(l_{2}),\cdots,\sigma_{1}(l_{\mathcal{N}})),\\
		\label{G-def}
		&\bG=(\bG_{i,j})_{\mathcal{N}\times \mathcal{N}},\quad \bG_{i,j}=\dfrac{1}{k_{i}-l_{j}}.	
	\end{align}
\end{subequations}
Therefore, the solution $\left( \mu,\nu\right)$ represented by \eqref{sym-FL-solu} can be rewritten as
\begin{subequations}
	\label{sym-FL-solu1}
	\begin{align}
		\mu=&\be_{\mathcal{N}}^{\st}\left( \bbD^{-1}+\bG\bbF\bG^{\st}\right)^{-1}\be_{\mathcal{N}},\\
		\nu=&\frac{-\mathrm{i}\be_{\mathcal{N}}^{\st}
\left( \bbF^{-1}+\bG^{\st}\bbD\bG\right)^{-1}\bK_{2}^{-1}
\be_{\mathcal{N}}}{1+\be_{\mathcal{N}}^{\st}\left( \bG^{\st}+\bbF^{-1}\bG^{-1}\bbD^{-1}\right)^{-1}\bK_{2}^{-1}\be_{\mathcal{N}}},
	\end{align}
\end{subequations}
where
\begin{align}
	\label{cal-PQ-def}
	\bbD=\bR_{1}\bS_{1},\quad\bbF=\bR_{2}\bS_{2}.
\end{align}
Functions \eqref{sym-FL-solu1} provide soliton solutions to
the discrete Mikhailov model \eqref{dcFL-1} through the
AKNS-type Cauchy matrix scheme.

\vskip 5pt
\noindent
\textbf{Multiple-pole solutions}. This type solutions
arise from the Jordan-block structure of matrices $\bK_1$ and $\bK_2$.
Here, we consider two forms of the matrices
$\bK_{1}$ and $\bK_{2}$.

\noindent$\bullet$\hspace{1em}\textit{Case-I}. We take $\bK_{1}$ as a Jordan-block matrix and
$\bK_{2}$ as a Jordan-block-like matrix
\begin{align}
\label{K12-Jor}
\bK_{1}=\bJ[k,1], \quad \bK_{2}=\bJ[l,-1], \quad \text{with} \quad k\neq l,
\end{align}
where
\begin{align}
\label{J-def}
		&\bJ[z,\delta]:=\begin{pmatrix}
			z&0&0&\cdots&0\\
			\delta&z&0&\cdots&0\\
			0&\delta&z&\cdots&0\\
			\vdots&\vdots&\vdots&\ddots&\vdots\\
			0&0&0&\cdots&z
\end{pmatrix}_{\mathcal{N}\times \mathcal{N}}.
\end{align}
Here we call $\bK_{2}$ a Jordan-block-like matrix since it has a similar
structure to the canonical Jordan-block form, while each element of the superdiagonal is $-1$.
In this situation, we have
\begin{subequations}
\label{JS-sym}
	\begin{align}
		&\br_{1}=\bF_{k}[\rho_{1}(k)]\bar{\be}_{\mathcal{N}},\quad \br_{2}=\bF_{-l}[\rho_{2}(l)]\bar{\be}_{\mathcal{N}}, \\
		&\bs_{1}=\bH_{k}[\sigma_{2}(k)]\bar{\be}_{\mathcal{N}},\quad
\bs_{2}=\bH_{-l}[\sigma_{1}(l)]\bar{\be}_{\mathcal{N}},\\
		&\bM_{1}=\bF_{k}[\rho_{1}(k)]\bar{\bG}\bH_{-l}[\sigma_{1}(l)],\quad
\bM_{2}=-\bF_{-l}[\rho_{2}(l)]\bar{\bG}\bH_{k}[\sigma_{2}(k)], \\
		&\bar{\bG}=(\bar{G}_{i,j})_{\mathcal{N}\times \mathcal{N}},\quad \bar{G}_{i,j}=\frac{(-1)^{i+j}
\mathrm{C}^{i-1}_{i+j-2}}{(k-l)^{i+j-1}},\quad \bar{\be}_{\mathcal{N}}=(\underbrace{1, 0, \cdots, 0}_{\mathcal{N}\text{-dimensional}})^{\st},
	\end{align}
\end{subequations}
where $\mathrm{C}^{i}_{j}=\frac{j!}{i!(j-i)!}$ for $j\geq i$ and
\begin{subequations}
	\label{FH-def}
	\begin{align}
\label{F-def}
		&\bF_{\delta z}[f(z)]=\begin{pmatrix}
			f(z)&0&0&\cdots&0\\
			\partial_{\delta z}f(z)&f(z)&0&\cdots&0\\
			\frac{\partial^{2}_{\delta z}f(z)}{2!}&\partial_{\delta z}f(z)&f(z)
&\cdots&0\\
			\vdots&\vdots&\vdots&\ddots&\vdots\\
			\frac{\partial^{\mathcal{N}-1}_{\delta z}f(z)}{(\mathcal{N}-1)!}
&\frac{\partial^{\mathcal{N}-2}_{\delta z}f(z)}{(\mathcal{N}-2)!}
&\frac{\partial^{\mathcal{N}-3}_{\delta z}f(z)}{(\mathcal{N}-3)!}&\cdots&f(z)
		\end{pmatrix}_{\mathcal{N}\times \mathcal{N}},\\
\label{H-def}
		&\bH_{\delta z}[f(z)]=
	\begin{pmatrix}
		\frac{\partial^{\mathcal{N}-1}_{\delta z}f(z)}{(\mathcal{N}-1)!} & \cdots 
& \frac{\partial^{2}_{\delta z}f(z)}{2!} & \partial_{\delta z}f(z) & f(z)\\
		\frac{\partial^{\mathcal{N}-2}_{\delta z}f(z)}{(\mathcal{N}-2)!} 
& \cdots & \partial_{\delta z}f(z) & f(z) & 0\\
		\frac{\partial^{\mathcal{N}-3}_{\delta z}f(z)}{(\mathcal{N}-3)!} & \cdots & f(z) & 0 & 0\\
		\vdots & \vdots & \vdots & \ddots &\ddots\\
		f(z) & \cdots & 0 & 0 & 0
	\end{pmatrix}_{\mathcal{\mathcal{N}}\times \mathcal{\mathcal{N}}}.
	\end{align}
\end{subequations}
Substituting \eqref{K12-Jor} and \eqref{JS-sym} into \eqref{sym-FL-solu} leads to a
$\mathcal{N}$-th order multiple-pole solution to the discrete Mikhailov model \eqref{dcFL-1}.

\begin{remark}
$\bF_{\delta z}[f(z)]$ in \eqref{F-def} is a lower triangular Toeplitz matrix commuting with a Jordan block of same order.
All such matrices compose a commutative set $\c{G}^{[\mathcal{N}]}$ with respect to matrix multiplication
and the subset
\[G^{[\mathcal{N}]}=\big \{\mathcal{C} \big |~\big. \mathcal{C}\in \c{G}^{[\mathcal{N}]},~|\mathcal{C}|\neq 0 \big\}\]
is an Abelian group. Such kind of matrices play useful roles in the expression of multiple-pole solutions.
For more properties of such matrices one can refer to \cite{Z-arxiv-2006,ZZSZ-RMP-2014}.
$\bH_{\delta z}[f(z)]$ in \eqref{H-def} is a skew triangular Hankel matrix, which is symmetric. It is easy to find that
$\bH_{\delta z}[f(z)]\bF_{\delta z}[f(z)]$ is a symmetric matrix, i.e.
$\bH_{\delta z}[f(z)]\bF_{\delta z}[f(z)]=\bF^{\st}_{\delta z}[f(z)]\bH_{\delta z}[f(z)]$.
\end{remark}

\noindent$\bullet$\hspace{1em}\textit{Case-II}. We take
\begin{align}
\bK_{1}=\bJ[k,1],\quad\bK^{\st}_{2}=\bJ[l,1].
\end{align}
In this case, we have
\begin{subequations}
	\begin{align}
		&\br_{1}=\bF_{k}[\rho_{1}(k)]\bar{\be}_{\mathcal{N}},\quad \br_{2}=\bH_{l}[\rho_{2}(l)]\bar{\be}_{\mathcal{N}},\\
		&\bs_{1}=\bH_{k}[\sigma_{2}(k)]\bar{\be}_{\mathcal{N}},\quad
\bs_{2}=\bF_{l}[\sigma_{1}(l)]\bar{\be}_{\mathcal{N}},\\
		&\bM_{1}=\bF_{k}[\rho_{1}(k)]\bG_{1}\bF^{\st}_{l}[\sigma_{1}(l)],
\quad\bM_{2}=\bH_{l}[\rho_{2}(l)]\bG_{2}\bH_{k}[\sigma_{2}(k)],\\
		&\bG_{1}=(g^{(1)}_{i,j})_{\mathcal{N}\times \mathcal{N}},\quad g^{(1)}_{i,j}=\frac{(-1)^{i-1}\mathrm{C}^{i-1}_{i+j-2}}{(k-l)^{i+j-1}},\\
		&\bG_{2}=(g^{(2)}_{i,j})_{\mathcal{N}\times \mathcal{N}},\quad g^{(2)}_{i,j}=\frac{(-1)^{i-1}\mathrm{C}^{i-1}_{i+j-2}}{(l-k)^{i+j-1}},
	\end{align}
\end{subequations}
which yield another $\mathcal{N}$-th order multiple pole solution to   \eqref{dcFL-1}.

\subsection{The KP-type scheme}\label{sec-3-2}

This case indicates $\bL\neq \bK$, $\bC=\bm 0$ and $\{\bS^{(i,j)}\}$
take the form \eqref{Sij-KP} where the involved elements are determined from \eqref{DES}.
To present solutions of \eqref{DES},
we denote $\br=(\br_{1},\br_{2})$ and $\bs=(\bs_{1},\bs_{2})$
with $\br_{\iota},  \bs_{\iota} \in\mathbb{C}^{N\times 1},~(\iota=1,2)$.
Taking these expressions into \eqref{DES} and writing $\bM=\bM_{1}+\bM_{2}$,
we arrive at two separated systems:
\begin{subequations}
	\label{asym-shift1}
	\begin{align}
     \label{asym-SE12}
	    & \bK\bM_{\iota}-\bM_{\iota}\bL=\br_{\iota}\bs^{\st}_{\iota}, \\
		\label{asym-shift-r1}
		& p\wt{\br}_{\iota}=p\br_{\iota}+(-1)^{\iota-1}\bK\br_{\iota},\quad q\wh{\br}_{\iota}=q\br_{\iota}+(-1)^{\iota-1}\bK^{-1}\br_{\iota}, \\
		\label{asym-shift-s1}
        & p\wt{\bs}_{\iota}+(-1)^{\iota-1}\bL^{\st}~\wt{\bs}_{\iota}=p\bs_{\iota},\quad
        q\wh{\bs}_{\iota}+(-1)^{\iota-1}(\bL^{-1})^{\st}~\wh{\bs}_{\iota}=q\bs_{\iota},
	\end{align}
\end{subequations}
with $\iota=1,2$,
which are the Cauchy matrix schemes for the scalar KP system \cite{HJN-book-2016,FZ-S-2022}.
Note that the notations $\br_{\iota},  \bs_{\iota}$ and $\bM_{\iota}$ ect.
are only used in this subsection and they
should NOT be mixed with those used in Sec.\ref{sec-3-1} in the AKNS-type scheme.

The master functions in this case are described as
\begin{align}
	\label{asym-S-def}
		\bS^{(i,j)}_{[\text{KP}]}=&\bs^{\st}\bL^{j}\bM^{-1}\bK^{i}\br
=\bs^{\st}\bL^{j}(\bM_{1}+\bM_{2})^{-1}\bK^{i}\br=\begin{pmatrix}
			s^{(i,j)}_{\text{[KP1]}}&s^{(i,j)}_{\text{[KP2]}} \\
			s^{(i,j)}_{\text{[KP3]}}&s^{(i,j)}_{\text{[KP4]}}
		\end{pmatrix} \nn \\
		= & \begin{pmatrix}		\bs^{\st}_{1}\bL^{j}(\bM_{1}+\bM_{2})^{-1}\bK^{i}\br_{1}
&\bs^{\st}_{1}\bL^{j}(\bM_{1}+\bM_{2})^{-1}\bK^{i}\br_{2}\\
\bs^{\st}_{2}\bL^{j}(\bM_{1}+\bM_{2})^{-1}\bK^{i}\br_{1}
&\bs^{\st}_{2}\bL^{j}(\bM_{1}+\bM_{2})^{-1}\bK^{i}\br_{2}
		\end{pmatrix},
\end{align}
and the matrix functions $\bu$ and $\bv$ defined in \eqref{uvw-def} read
\begin{align}
	\label{asmy-uv-def}
	\bu=\bS^{(0,0)}_{[\text{KP}]}=\bs^{\st}\bM^{-1}\br,\quad
\bv=\bI-\bS^{(-1,0)}_{\text{[KP]}}=\bI-\bs^{\st}\bM^{-1}\bK^{-1}\br.
\end{align}
Although
\begin{align}
	\label{asym-det-v}
	\left| \bv\right|=\left|\bI-\bs^{\st}\bM^{-1}\bK^{-1}\br \right|
=\left| \bL\right|/\left| \bK\right|,
\end{align}
which is a constant (maybe   not 1),  the normalization \eqref{trans1}
 does not change the expressions of solutions:
\begin{align}
	\label{asym-FL-solu}
	\left( \mu,\nu\right) =& \left( u_{3},\mathrm{i}\frac{V_{2}}{V_{4}}\right) 
=\left( u_{3},\mathrm{i}\frac{v_{2}}{v_{4}}\right) \nn \\
       =& \left(\bs^{\st}_{2}(\bM_{1}+\bM_{2})^{-1}\br_{1},
       \frac{-\mathrm{i}\bs^{\st}_{1}
       (\bM_{1}+\bM_{2})^{-1}\bK^{-1}\br_{2}}{1-\bs^{\st}_{2}(\bM_{1}
       +\bM_{2})^{-1}\bK^{-1}\br_{2}}\right).
\end{align}
Therefore the normalization \eqref{trans1} is valid.

Then, by solving \eqref{asym-shift1} we may obtain explicit form for
\eqref{asym-FL-solu} to describe multi-soliton solutions and multiple-pole solutions to
the discrete Mikhailov model \eqref{dcFL-1}.
Below we list out formulae for  the involved elements of these solutions.

\vskip 5pt

\noindent
\textbf{Soliton solutions}.
\begin{subequations}
	\label{asym-SE-solu}
	\begin{align}
	& \bK=\text{diag}(k_{1},k_{2},\cdots,k_{N}),\quad\bL=\text{diag}(l_{1},l_{2},\cdots,l_{N}),\\
	\label{asym-M1M2-decom}
	& \bM_{\iota}=\bR_{\iota}\bG\bS_{\iota}, \quad \br_{\iota}=\bR_{\iota}\be_{N}, \quad
\bs_{\iota}=\bS_{\iota} \be_{N},\quad (\iota=1,2),
	\end{align}
\end{subequations}
with $k_i\neq l_j,~(i,j=1,2,\ldots,N$), in which
\begin{subequations}
	\label{asym-RGS-def}
	\begin{align}
		\bR_{\iota}=&\text{diag}(\rho_{\iota}(k_{1}),\rho_{\iota}(k_{2}),\cdots,\rho_{\iota}(k_{N})), \quad
        \bS_{\iota}=\text{diag}(\sigma_{\iota}(l_{1}),\sigma_{\iota}(l_{2}),\cdots,\sigma_{\iota}(l_{N})),\\
		\bG=&(\bG_{i,j})_{N\times N},\quad\bG_{i,j}=\frac{1}{k_{i}-l_{j}},\quad\be_{N}
=(\underbrace{1, 1, \cdots, 1}_{N\text{-dimensional}})^{\st},
	\end{align}
\end{subequations}
where $\rho_{\iota},~\sigma_{\iota}~(\iota=1,2)$ are defined in \eqref{plane-def}. We denote $\bR=\bR_{1}^{-1}\bR_{2}, \bS=\bS_{1}^{-1}\bS_{2}$
and rewrite solution \eqref{asym-FL-solu} as
\begin{subequations}
	\label{asym-FL-solu1}
	\begin{align}
		\mu=&\be_{N}^{\st}\left( \bG\bS^{-1}+\bR\bG\right) ^{-1}\be_{N},\\
		\nu=&-\frac{\mathrm{i}\be_{N}^{\st}\left( \bR^{-1}\bG+\bG\bS\right)^{-1}\bK^{-1}\be_{N}}{1-\be_{N}^{\st}
\left( \bR^{-1}\bG\bS^{-1}+\bG\right) ^{-1}\bK^{-1}\be_{N}}.
	\end{align}
\end{subequations}

\vskip 5pt
\noindent
\textbf{Multiple-pole solutions}.
\begin{subequations}
	\begin{align}
		&\bK=\bJ[k,1],\quad\bL^{\st}=\bJ[l,1],\\
		&\br_{\iota}=\bF_{k}[\rho_{\iota}(k)]\bar{\be}_{N},\quad
\bs_{\iota}=\bF_{l}[\sigma_{\iota}(l)]\bar{\be}_{N}, \quad \bar{\be}_{N}
=(\underbrace{1, 0, \cdots, 0}_{N\text{-dimensional}})^{\st}, \\
		&\bM=\bM_{1}+\bM_{2},\quad\bM_{\iota}
=\bF_{k}[\rho_{\iota}(k)]\bar{\bG}\bF^{\st}_{l}[\sigma_{\iota}(l)],
\quad \iota=1,2,\\
		&\bar{\bG}=(\bar{G}_{i,j})_{N\times N},\quad \bar{G}_{i,j}=\frac{(-1)^{i-1}\mathrm{C}^{i-1}_{i+j-2}}{(k-l)^{i+j-1}},
	\end{align}
\end{subequations}
where the matrices $\bJ$ and $\bF$ are defined by \eqref{J-def} and \eqref{F-def}, respectively.
By these entries, \eqref{asym-FL-solu} yields a
$N$-th order multiple-pole solution for  \eqref{dcFL-1}.

\section{Continuum limits}\label{sec-4}

We explore in this section continuum limits of the discrete Mikhailov model \eqref{dcFL-1}
by shrinking the lattice grid to a continuous set of values corresponding to spatial and temporal coordinates.
From the  system \eqref{dcFL-1}, through one step straight continuum limit, we obtain
a semi-discrete Mikhailov model (see \eqref{semi-continuous FL}),
i.e., the differential-difference Mikhailov model with one discrete and one continuous independent variables.
Subsequently, Cauchy matrix schemes of the AKNS-type and the KP-type will be
presented for the semi-discrete Mikhailov model.
The second step continuum limit leading to the continuous Mikhailov model \eqref{MM} will be
performed in this section as well.

\subsection{Semi-discrete Mikhailov model}\label{sec-4-1}

In the first step we let both $q$ and $m$ tend to infinity but keep $m/q$ finite.
Introduce
\begin{equation}\label{xmq}
x=\frac{m}{q},
\end{equation}
where $1/q$ serves as the grid parameter in the discretization of $x$.
With this setting
we reinterpret the dependent variables $\mu$ and $\nu$ as
\begin{align}
	\label{ctn-uv}
	\mu(n,m)=\mu(n,x),\quad\nu(n,m)=\nu(n,x)
\end{align}
without making confusions,
while the shifted variables in $m$-direction give rise to
\begin{subequations}
	\label{Taylor-expan-uv}
	\begin{align}
		\label{Taylor-expan-u}
		\wh{\mu}=&\mu(n,x+1/q)=\mu+\mu_{x}/q+\mu_{xx}/(2q^{2})+\cdots,\\
		\label{Taylor-expan-wu}
		\wh{\wt{\mu}}=&\mu(n+1,x+1/q)=\wt{\mu}+\wt{\mu}_{x}/q+\wt{\mu}_{xx}/(2q^{2})+\cdots,
	\end{align}
\end{subequations}
and a similar formula  for  $\wh{\wt{\nu}}$.
Inserting them into the system \eqref{dcFL-1} and then
from the leading term (in terms of $1/q$) we obtain a semi-discrete system:
\begin{subequations}
	\label{semi-continuous FL}
\begin{align}
	\label{semi-continuous FL1}
	&(\wt{\nu}+\nu)(\wt{\mu}+\mu)(\wt{\mu}_{x}+\mu_{x})
-\mathrm{i}[(\mu_{x}-\wt{\mu}_{x}) P+4(\wt{\mu}+\mu)] =0,\\
	\label{semi-continuous FL2}
	&(\wt{\nu}^{2}-\nu^{2})(\wt{\mu}_{x}-\mu_{x})-2(\wt{\mu}
+\mu)(\wt{\nu}_{x}\wt{\nu}+\nu_{x}\nu)
-\mathrm{i}[4(\wt{\nu}+\nu)-(\wt{\nu}_{x}-\nu_{x})P-(\wt{\nu}-\nu)P_{x}]=0,
\end{align}
where
\begin{align}
	P:=P(n,x)=[4p^2-(\mu+\wt{\mu})^{2}(\nu+\wt{\nu})^{2}]^{\frac{1}{2}}.
\end{align}
\end{subequations}
This is the semi-discrete Mikhailov model.

In the above limit, the discrete plane wave factors in \eqref{plane-def} give rise to
\begin{subequations}
	\label{semi-pwf}
\begin{align}
	&\rho_{\iota}(z)=\left(1+ (-1)^{\iota-1}\frac{z}{p}\right)^{n}
e^{(-1)^{\iota-1}z^{-1}x}\rho^{0}_{\iota}(z), \\
	&\sigma_{\iota}(z)=\left(1+ (-1)^{\iota-1}\frac{z}{p}\right)^{-n}
e^{(-1)^{\iota}z^{-1}x}\sigma^{0}_{\iota}(z),
\end{align}
\end{subequations}
with $\iota=1,2$,
which now serve for the semi-discrete system \eqref{semi-continuous FL}.
Moreover, solutions for semi-discrete Mikhailov model can be provided from
Cauchy matrix schemes as well.

\begin{prop}\label{Prop-4-1}
Formulae \eqref{sym-FL-solu} and \eqref{asym-FL-solu} provide solutions to
the semi-discrete Mikhailov model \eqref{semi-continuous FL}, respectively,
through the Cauchy matrix scheme of the AKNS-type:
\begin{subequations}
	\label{sd-DES-sym-ent}
	\begin{align}
        \label{sd-sym-SE12}
	    & \bK_{1}\bM_{1}-\bM_{1}\bK_{2}=\br_{1}\bs^{\st}_{2},\quad \bK_{2}\bM_{2}-\bM_{2}\bK_{1}=\br_{2}\bs^{\st}_{1}, \\
		\label{sd-sym-shift-r1}
		&  p\wt{\br}_{\iota}=p\br_{\iota}+(-1)^{\iota-1}\bK_{\iota}\br_{\iota},\quad \br_{\iota,x}=(-1)^{\iota-1}\bK^{-1}_{\iota}\br_{\iota}, \\
		\label{sd-sym-shift-s1}
		& p\wt{\bs}_{\iota}+(-1)^{\iota}\bK^{\st}_{\iota}~\wt{\bs}_{\iota}=p\bs_{\iota},\quad \bs_{\iota,x}=(-1)^{\iota-1}(\bK^{-1}_{\iota})^{\st}\bs_{\iota},
	\end{align}
\end{subequations}
and through the Cauchy matrix scheme of the KP-type:
\begin{subequations}
	\label{sd-asym-shift1}
	\begin{align}
     \label{sd-asym-SE12}
	    & \bK\bM_{\iota}-\bM_{\iota}\bL=\br_{\iota}\bs^{\st}_{\iota},  \\
		\label{sd-asym-shift-r1}
		& p\wt{\br}_{\iota}=p\br_{\iota}+(-1)^{\iota-1}\bK\br_{\iota},\quad \br_{\iota,x}=(-1)^{\iota-1}\bK^{-1}\br_{\iota}, \\
		\label{sd-asym-shift-s1}
        &  p\wt{\bs}_{\iota}+(-1)^{\iota-1}\bL^{\st}~\wt{\bs}_{\iota}=p\bs_{\iota}, \quad
        \bs_{\iota,x}=(-1)^{\iota}(\bL^{-1})^{\st}~\bs_{\iota},
	\end{align}
\end{subequations}
with $\iota=1,2$.
\end{prop}

\subsection{Continuous Mikhailov model}

To implement the other straight continuum limit, we set
\begin{align}
	\label{lim-n}
	n\rightarrow\infty,\quad p\rightarrow\infty,\quad\text{while}\quad t:=n/p  \quad\text{finite},
\end{align}
where $t$ is the temporal continuous coordinate.
We also set
\[\mu(n,x)=\mu(t,x),\quad \nu(n,x)=\nu(t,x),\]
and as a consequence, we have
\begin{subequations}
	\label{Taylor-expan-uv-wt}
	\begin{align}
		\label{Taylor-expan-u-wt}
		\wt{\mu}=&\mu(t+1/p,x)=\mu+\mu_{t}/p+\mu_{tt}/(2p^{2})+\cdots,\\
		\label{Taylor-expan-v-wt}
		\wt{\nu}=&\nu(t+1/p,x)=\nu+\nu_{t}/p+\nu_{tt}/(2p^{2})+\cdots.
	\end{align}
\end{subequations}
Inserting them into the semi-discrete system \eqref{semi-continuous FL},
the leading term recovers a coupled system
\begin{subequations}
	\label{KN(-1)-eq-system}
	\begin{align}
		\label{KN(-1)-eq1}
		&\mu_{xt}-4\mu-4\mathrm{i}\mu\nu\mu_{x}=0,\\
		\label{KN(-1)-eq2}
		&\nu_{xt}-4\nu+4\mathrm{i}\mu\nu\nu_{x}=0,
	\end{align}
\end{subequations}
which (after a coordinate scaling $(t,x) \rightarrow (t/2,-x/2)$)
is nothing but the Mikhailov model \eqref{MM}.

After the  continuum limit \eqref{lim-n}, the plane wave factors are
\begin{subequations}
	\label{ctn-plane-def}
	\begin{align}
		&\rho_{\iota}(z)=e^{(-1)^{\iota-1}(zt+z^{-1} x)}\rho^{0}_{\iota}(z),\\
		&\sigma_{\iota}(z)=e^{(-1)^{\iota}(zt+z^{-1} x)}\sigma^{0}_{\iota}(z),
	\end{align}
\end{subequations}
with $\iota=1,2$.
And moreover, solutions for the Mikhailov model \eqref{KN(-1)-eq-system} can be provided from
Cauchy matrix schemes as well,
which correspond to the results obtained in \cite{LLZ-SAPM-2025}.

\begin{prop}
Formulae \eqref{sym-FL-solu} and \eqref{asym-FL-solu} provide solutions to
the continuous Mikhailov model \eqref{KN(-1)-eq-system}, respectively,
through the Cauchy matrix scheme of the AKNS-type:
\begin{subequations}
	\label{c-DES-sym-ent}
	\begin{align}
        \label{c-sym-SE12}
	    & \bK_{1}\bM_{1}-\bM_{1}\bK_{2}=\br_{1}\bs^{\st}_{2},\quad \bK_{2}\bM_{2}-\bM_{2}\bK_{1}=\br_{2}\bs^{\st}_{1}, \\
		\label{c-sym-shift-r1}
		& \br_{\iota,t}=(-1)^{\iota-1}\bK_{\iota}\br_{\iota},\quad \br_{\iota,x}=(-1)^{\iota-1}\bK^{-1}_{\iota}\br_{\iota}, \\
		\label{c-sym-shift-s1}
		& \bs_{\iota,t}=(-1)^{\iota-1}\bK^{\st}_{\iota}\bs_{\iota},\quad \bs_{\iota,x}=(-1)^{\iota-1}(\bK^{-1}_{\iota})^{\st}~\bs_{\iota},
	\end{align}
\end{subequations}
through the Cauchy matrix scheme of the KP-type:
\begin{subequations}
	\label{c-asym-shift1}
	\begin{align}
     \label{c-asym-SE12}
	    & \bK\bM_{\iota}-\bM_{\iota}\bL=\br_{\iota}\bs^{\st}_{\iota},  \\
		\label{c-asym-shift-r1}
		& \br_{\iota,t}=(-1)^{\iota-1}\bK\br_{\iota},\quad \br_{\iota,x}=(-1)^{\iota-1}\bK^{-1}\br_{\iota}, \\
		\label{c-asym-shift-s1}
        & \bs_{\iota,t}=(-1)^{\iota}\bL^{\st}\bs_{\iota},\quad
        \bs_{\iota,x}=(-1)^{\iota}(\bL^{-1})^{\st}\bs_{\iota},
	\end{align}
\end{subequations}
with $\iota=1,2$.
 \end{prop}

\section{Concluding remarks}\label{sec-5}

In this paper we have derived the discrete Mikhailov model \eqref{dcFL-1}
by means of the Cauchy matrix approach.
The key step in the discretization is to find a pair of discrete Miura transformations,
i.e. \eqref{Miura-h} and \eqref{Miura-t}.
Among the two equations in the discrete Mikhailov model \eqref{dcFL-1},
one equation, \eqref{dcFL-1a}, results from equation \eqref{nAKNS2} by means of the Miura transformations,
the other, \eqref{dcFL-1b},  is from the compatibility of the two Miura transformations.
This inherits the continuous  Cauchy matrix structure (cf.\cite{LLZ-SAPM-2025})
except the continuous Miura transformation contains a single equation ((2.13) in \cite{LLZ-SAPM-2025}).
Apart from equations, we also constructed explicit solutions, including solitons
and multiple-pole solutions, from two Cauchy matrix schemes,
namely, the AKNS-type and the KP-type.
By straight continuum limits we recovered the semi-discrete and continuous Mikhailov models
together with their Cauchy matrix structures and solutions.
It is notable that the continuum limits show that the discretization is direct,
i.e. the discrete variables $n$ and $m$  directly correspond to $t$ and $x$, respectively.

Let us address several problems related to the approach and results of the present paper.
First, in this paper we not only presented a pair of discrete Miura transformations,
but also we have showed a way how to construct them.
This is useful in discretizing other integrable equations where Miura transformations are involved.
For example, it is known that there is a Miura transformation
to connect the unreduced NLS system and the unreduced
Gerdjikov-Ivanov (GI) derivative NLS system \cite{KK-IMRN-2004}.
This indicates the GI  system might be discretized if proper discrete Miura transformations could be found.
It then possible to formulate the Cauchy matrix structures
for all the discrete derivative NLS type systems.
The second problem is about possible reductions of the discrete Mikhailov model \eqref{dcFL-1} or
\eqref{dcFL-2}.
In Appendix we collect some fully discrete equations of the unreduced NLS and derivative NLS systems.
Neither of them admits one-field reduction.
So far we do not find any one-field reduction for \eqref{dcFL-1} either.
However, in \cite{NCQ-PLA-1983} the massive Thirring model has been discretized.
Whether the discrete Mikhailov model \eqref{dcFL-1} or
\eqref{dcFL-2} can be reduced to the discrete massive Thirring model will be investigated later.
Besides, what we have not touched upon  is the Lax integrability of the discrete Mikhailov system \eqref{dcFL-1}.
In fact, in the Cauchy matrix, there is an object $\bu^{(i)}=(\bC+\bM)^{-1}\bK^{i}\br$
can be defined along with $\bS^{(i,j)}$.
Note that in this paper  $\bu^{(i)}$ is a $N\times 2$ matrix rather than a vector.
Such $\bu^{(i)}$ will satisfy a set of linear relations which can be used to
construct Lax pairs for some equations involving $\bS^{(i,j)}$.
As examples one can refer to \cite{CWZ-TMP-2023} and Chapter 9 of \cite{HJN-book-2016}
where $\bu^{(i)}$ is a vector.
However, for the discrete Mikhailov system \eqref{dcFL-1}, the involved variables are
$u_3$ and $v_2/v_4$,  and $\bu^{(i)}$ is no longer a vector,
one needs to develop special approach to construct Lax pairs by using $\bu^{(i)}$.
In addition, for the discrete equations presented in Appendix,
their Lax pairs can be constructed from deforming Darboux transformations
of the related continuous spectral problems,
which implies the multidimensional consistency (see \cite{HJN-book-2016}) behind them.
For the discrete Mikhailov system,
how to describe its multidimensional consistency (with other system)
and whether its Lax integrability can de described from this way will be also investigated in the future.

\vskip 20pt
\subsection*{Data availability statement}

Data sharing not applicable to this article as no datasets were generated or analysed during the current study.

\subsection*{Conflict of interest statement }

The authors declare that they have no conflict of interest.

\vskip 20pt
\subsection*{Acknowledgments}

This project is supported by the National Natural Science Foundation of China (Grant Nos. 12071432 and 12271334)
and the Natural Science Foundation of Zhejiang Province (Grant No. LZ24A010007).

\begin{appendix}

\titleformat{\section}{\Large\bfseries }{Appendix \thesection }{0.5em}{ }{}

\section{Discrete equations of the unreduced NLS type}\label{sec-app}

Here we collect some fully discrete two-field integrable equations.
Apart from the discrete AKNS$(-1)$ system \eqref{nAKNS},
there is a positive order discrete AKNS  system:
\begin{subequations}\label{dNLS}
\begin{align}
& (\wt{u}-\wh{u})(u\wh{\wt{v}}+1)+(p-q)u=0,\label{2.4a}\\
& (\wt{v}-\wh{v})(u\wh{\wt{v}}+1)-(p-q)\wh{\wt{v}}=0,\label{2.4b}
\end{align}
\end{subequations}
where $p$ and $q$ are spacing parameters.
In continuum limits it corresponds to the second order AKNS system (unreduced NLS system) (e.g. \cite{CXZ-proc-2020})
\begin{subequations}\label{cNLS}
\begin{align}
&  u_t-u_{xx}+2u^2v=0,\\
&  v_t+v_{xx}-2uv^2=0.
\end{align}
\end{subequations}
The discrete system \eqref{dNLS} was first derived in \cite{K-PLA-1982} in 1982 by Konopelchenko
as a superposition formula of the B\"acklund transformation of \eqref{cNLS},
and later by Date, Jimbo and Miwa in \cite{DJM-JPSJ-1983} in 1983.
It is also understood as a result of the compatibility of two Darboux transformations of the
AKNS spectral problem \cite{CXZ-proc-2020}.

Corresponding to the continuous KN system (unreduced derivative NLS system)
\begin{subequations}\label{NLS-KN}
\begin{align}
& u_t-u_{xx}+2(u^2v)_x=0,\\
& v_t+v_{xx}+2(uv^2)_x=0,
\end{align}
\end{subequations}
there is a discretization \cite{XCZ-JPA-2022}
\begin{subequations}\label{dKN}
\begin{align}
&(1+(1-4u\wt{v})^{\frac{1}{2}})(\wt{u}+p^2u)
-(1+(1-4u\wh{v})^{\frac{1}{2}}) (\wh{u}+q^2u)-2(\wt{u}\wt{v}-\wh{u}\,\wh{v})u=0, \\
& (1+(1-4\wt{u}\wh{\wt{v}})^{\frac{1}{2}})(\wt{v}+q^2\wh{\wt{v}})
- (1+(1-4\wh{u}\wh{\wt{v}})^{\frac{1}{2}})(\wh{v}+p^2\wh{\wt{v}})
-2(\wt{u}\wt{v}-\wh{u}\,\wh{v})\wh{\wt{v}}=0,
\end{align}
\end{subequations}
which is obtained from the compatibility of two Darboux transformations of the
KN spectral problem \cite{CXZ-proc-2020}.

The third one is a discretization of the unreduced Chen-Lee-Liu system:
\begin{subequations}\label{eq:0.000}
\begin{align}
& u_{t}+u_{xx}+2uvu_{x}=0,\label{eq:0.000a} \\
& v_{t}-v_{xx}+2uvv_{x}=0.\label{eq:0.000b}
\end{align}
\end{subequations}
The discrete system reads
\begin{subequations}\label{eq:1.1}
\begin{align}
& q^{2}\wh{u}-p^{2}\wt{u}
+u(p^{2}-q^{2})+u\wh{\wt{v}}(\wh{u}-\wt{u})=0,\label{eq:1.1a} \\
& p^{2}\wh{v}-q^{2}\wt{v}
+\wh{\wt{v}}(q^{2}-p^{2})+u\wh{\wt{v}}(\wh{v}-\wt{v})=0.\label{eq:1.1b}
\end{align}
\end{subequations}
It was presented by Date, Jimbo and Miwa in \cite{DJM-JPSJ-1983}.
It is also understood as  the compatibility of two Darboux transformations of the
Chen-Lee-Liu spectral problem \cite{LZZ-SAPM-2024,XYLZ-PD-2025}.

%\begin{align*}
%	& (\mathrm{i}Q(\nu+\wh{\nu})+(u_3-\wh{u}_3)(\nu^2+\wh{\nu}^2))(\mathrm{i}P+2u_3(\nu-\wt{\nu}))
%-(\mathrm{i}Q+(u_3-\wh{u}_3)(\nu+\wh{\nu}))\\
%& \quad \cdot (\mathrm{i}P(\nu-\wt{\nu})+2u_3(\nu^2+\wt{\nu}^2))-(\mathrm{i}\wt{Q}
%+(\wt{u}_3-\wh{\wt{u}}_3)(\wt{\nu}+\wh{\wt{\nu}}))
%(\mathrm{i}\wh{P}(\wh{\nu}-\wh{\wt{\nu}})+2\wh{u}_3(\wh{\nu}^2+\wh{\wt{\nu}}^2)) \\
%& \quad +(-\mathrm{i}\wh{P}+2\wh{u}_3(\wh{\wt{\nu}}-\wh{\nu}))(\mathrm{i}\wt{Q}(\wt{\nu}
%+\wh{\wt{\nu}})+(\wt{u}_3-\wh{\wt{u}}_3)(\wt{\nu}^2+\wh{\wt{\nu}}^2))=0.
%\end{align*}

\end{appendix}

\vskip 20pt

{\small

\end{document}